\documentclass{article}
\usepackage{spconf,amsmath,graphicx,supertabular,booktabs,dsfont, setspace}
\usepackage{xcolor}

\title{AttentionAnatomy: A unified framework for whole-body Organs at Risk Segmentation using multiple partially annotated datasets}
\name{Shanlin Sun$^{\star\mathsection}$, Yang Liu$^{\dagger\mathsection}$\thanks{$^{\mathsection}$ authors contributed equally}, Narisu Bai$^{\star}$, Hao Tang$^{\dagger}$, Xuming Chen$^{\ddagger}$, Qian Huang$^{\ddagger}$, Yong Liu$^{\ddagger}$, Xiaohui Xie$^{\dagger}$}

\address{$^{\star}$ DeepVoxel Inc, Irvine, CA, USA  \\$^{\ddagger}$ Department of Radiation Oncology, Shanghai General Hospital, \\ Shanghai Jiao Tong University School of Medicine, Shanghai, China \\$^{\dagger}$ Department of Computer Science, University of California, Irvine, CA, USA}

\begin{document}

%
\maketitle
\begin{abstract}

Organs-at-risk (OAR) delineation in computed tomography (CT) is an important step in Radiation Therapy (RT) planning. Recently, deep learning based methods for OAR delineation have been proposed and applied in clinical practice for separate regions of the human body (head and neck, thorax, and abdomen). However, there are few researches regarding the end-to-end whole-body OARs delineation because the existing datasets are mostly partially or incompletely annotated for such task. In this paper, our proposed end-to-end convolutional neural network model, called \textbf{AttentionAnatomy}, can be jointly trained with three partially annotated datasets, segmenting OARs from whole body. Our main contributions are: 1) an attention module implicitly guided by body region label to modulate the segmentation branch output; 2) a prediction re-calibration operation, exploiting prior information of the input images, to handle partial-annotation(HPA) problem; 3) a new hybrid loss function combining batch Dice loss and spatially balanced focal loss to alleviate the organ size imbalance problem. Experimental results of our proposed framework presented significant improvements in both Sørensen-Dice coefficient (DSC) and 95\% Hausdorff distance compared to the baseline model.

\end{abstract}
\section{Introduction}
\label{sec:intro}

Radiation Therapy(RT) is an important curative treatment for multiple types of cancers. A key step in RT planning is to accurately delineate all OARs in CT images. Recently, Deep Convolutional Neural Networks(DCNNs) methods have been successfully applied to different medical image segmentation tasks\cite{milletari2016v,ronneberger2015u}, including OAR delineation \cite{tang2019clinically, zhu2019anatomynet}. However, these methods are proposed for delineating OARs in only part of the human body, e.g. head and neck (HaN), throax and abdomen. A unified model for whole-body OAR delineation has great clinical implication, but only few researches focus on this topic.

The most innegligible reason is about data. As for the whole-body OARs delineation taks, the existing datasets are mostly partially labelled for three different parts of human body (head and neck, thorax, and abdomen). This poses great challenges in training an end-to-end deep learning model for whole-body delineation. For instance, datasets annotated for HaN delineation may contain CT scans including thorax region but only have OARs in the HaN annotated. If the unannotated thoracic anatomies are treated as background, the model may have difficulty learning the contradictory representation.

There are also three main challenges for this task. First, a naive/brute force approach, which first classifies the CT scan into three regions and then uses different segmentation models for different regions, may have systematic errors. Misclassification of different parts of human body will significantly affect the segmentation quality, which largely offsets the gain from automatic delineation in clinical practice. Second, current state-of-the-arts OARs delineation methods\cite{tang2019clinically, zhu2019anatomynet} use 3D convolutions and require whole volume CT image as input, which lacks scalability when applied to whole body because of memory constraints. Third, the imbalance of volume sizes of different OARs.

In this paper, we propose an end-to-end 2.5D DCNN framework, named \textbf{AttentionAnatomy}, to address the aforementioned challenges. AttentionAnatomy preserves the encoder-decoder structure of U-Net and has two branches: a CT region classification branch and an OAR segmentation branch. The CT region classification branch outputs a region predication as well as an attention vector of 33 elements, representing an inference of possible combination of OARs in current image. The OAR segmentation branch then uses this attention vector to modulate the final output mask. We further propose a re-calibration mechanism to tackle the partial-annotation problem, and a hybrid loss function consisting of batch dice loss and spatially balanced focal loss to cope with the extreme class imbalance. Experimental results showed AttentionAnatomy achieved a significant increase in DSC and drop in 95\% Hausdorff distance compared to the baseline model.

\section{Materials and Methods}
\label{sec:mm}

\subsection{Data}
\label{ssec:data}



We used three in-house datasets (Head and Neck(HaN), Thorax and Abdomen) in our study and each contains 41, 43 and 45 CT scans respectively. A total of 33 OARs were delineated by a radiation oncologist with more than 10 years of experience. We randomly split three datasets into 36, 37 and 39 for training and 5, 6 and 6 for testing. This leads to a total number of 112 CT images in the training set and 17 the test set.

Each CT scan is manually assigned one of the five classification labels: head, upper chest, chest, upper abdomen and abdomen.
And each slice of the CT image has resolution $512\times 512$, and we center crop a region of $320\times 320$ for faster training. We stack 5 continuous slices from the CT scan along the channel and feed this tensor into the proposed model.  The proposed model then outputs 34 2D binary masks, corresponding to the segmentation result of the input center slice, one for each OAR or background.

\subsection{Network architecture}
\label{ssec:net-arch}

\begin{figure}[htb]
\centering
\centerline{\includegraphics[width=8cm]{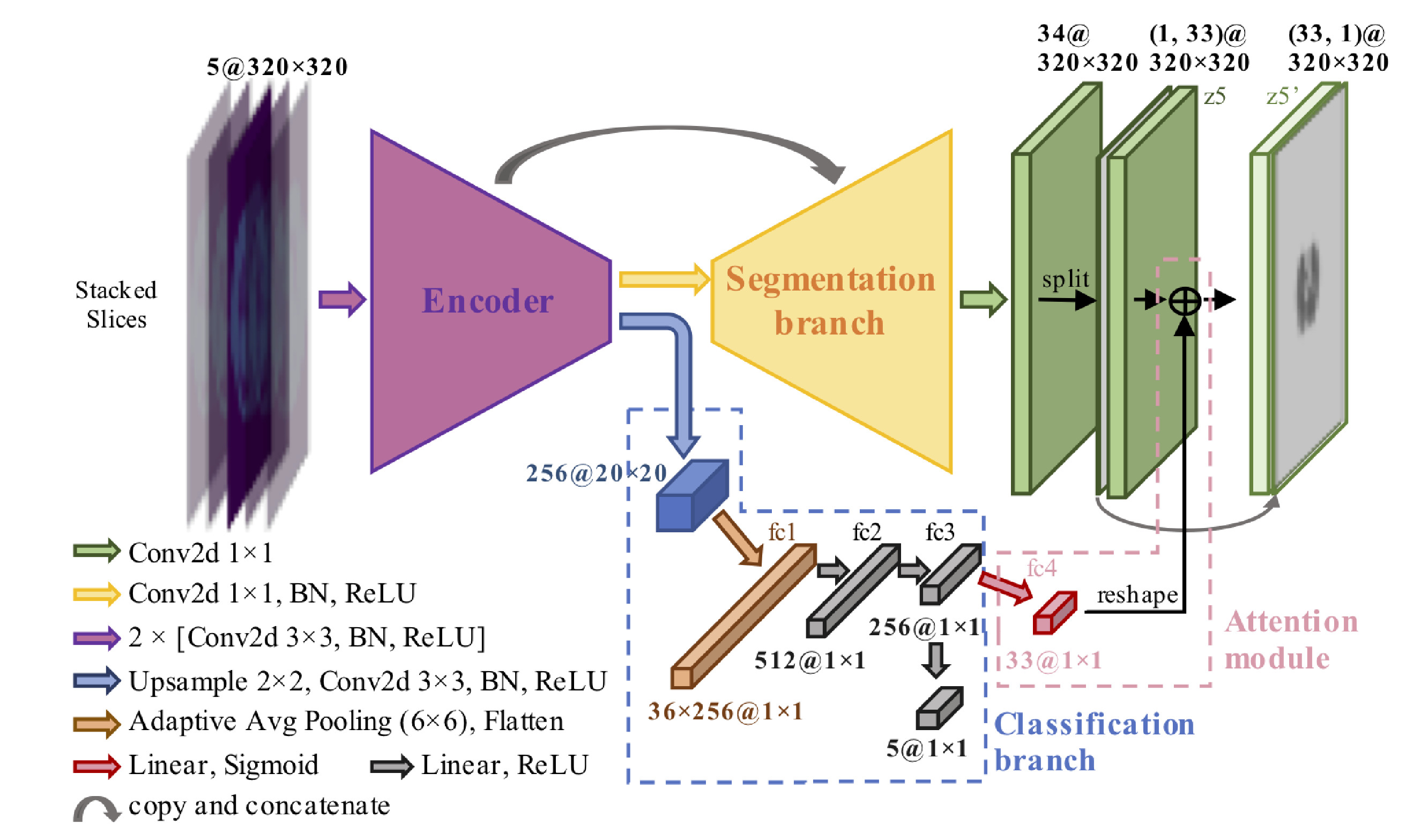}}
\caption{Overview of the model.}
\vspace{-2em}
\label{fig:model}
\end{figure}

Fig.\ref{fig:model} describes the architecture of AttentionAnatomy. The choice of encoder and decoder branches is flexible and not limited any particular implementations. We chose the standard residual U-Net in this work. $S$, $H$, $W $ are the number of slices, height and weight of the input images and $C$ is the channel number of the predicted segmentation. In our work, $S$, $H$, $W $, $C$ are 5, 320, 320 and 34 respectively.

\subsubsection{classification branch}
\label{sssec:classify}
The details of classification branch are shown in Fig.\ref{fig:model}, whose output is a scan-wise prediction to indicate which region one scan most probably belongs to. The principal purpose for designing such a classification branch is not to decide the type of scan during inference, but to enable the feature maps of classification branch to represent some general spatial information of the input scans. More specifically, we expect these feature map to help identify which OARs exist in the input scans.

\subsubsection{attention module}
\label{sssec:attention}
Attention module, connecting classification branch and segmentation branch as shown in Fig.\ref{fig:model}, aims to modulate the probability prediction of segmentation branch. It is designed to suppress predictions of OARs which do not exist in the input scans at the same time, it should assign bigger weights on OARs which exist in the input scans. For class c, $z^4 \in \mathds{R}^{C-1}$  is the output of 'fc4' layer as shown in Fig.\ref{fig:model}, and $z_n^5 \in \mathds{R}^{C-1} $ is n-th voxel of the feature map $z^5 \in \mathds{R}^{(C-1) \times H \times W}$and $z_n^{5'} \in \mathds{R}^{C-1} $ is is n-th voxel of the feature map $z^{5'} \in \mathds{R}^{(C-1) \times H \times W}$(both shown in Fig.\ref{fig:model}). The attention module works as
\vspace{-0.5mm}
\begin{equation}
z_n^{ 5^{'}}(c)=\ln(z^4(c)+\epsilon)+z_n^5(c),\:\:\:\:\:c = 0,1,...,C-2
\end{equation}

At the end, using softmax activation, for each voxel n we get the predicted probability $p_n(c) = softmax(z_n^{ 5^{'}}) \approx z^4 \bigotimes softmax(z_n^5 )$. Here, we decoupled attention module with last layer of classification branch, because the classification task only cares about the distinctive features among different regions while the OARs of interest in different types of regions are not mutually exclusive. For example, spinal cord lies across the upper body, thus spinal cord hardly matters in scan classification but the segmentation branch should pay attention to it for most scans. To avoid model from sticking at the sub-optimal point resulting from the mismatched goals of classification and segmentation, we designed the architecture that attention module and classification branch share features except for their last layers.

\subsection{Loss function}
\label{ssec:loss-fn}
For the 2.5D model, volume size imbalance problem are introduced from two perspectives---spatial size in x-y plane and length in z axis. For example, spinal cord could be seen as a small anatomy structure in one single CT scan, but it could seen as a 'long' structure because it tends to occupy many scans in a CT volume. In contrast, sublingual gland is not a very small organ spatially, but is so 'short' that only occupies two or three scans in the CT volume. We employ a hybrid segmentation loss combining batch dice loss and spatially balance focal loss to alleviate the volume size imbalance problem. In terms of classification branch, we simply apply cross entropy loss. Thus, the total loss can be expressed as

\begin{equation}
\mathcal{L} =  \alpha\mathcal\mathcal{L}_{batch\_dice} + \beta\mathcal{L}_{sb\_focal} + \gamma\mathcal{L}_{ce}
\end{equation}

where $\alpha$, $\beta$ and $\gamma$ are trade-offs among batch dice loss $\mathcal{L}_{batch\_dice}$, spatially balanced focal loss $\mathcal{L}_{sb\_focal}$ and region classification cross entropy loss $\mathcal{L}_{ce}$.

In what follows, \(p_{ij}(c)\) is the predicted probability for voxel i at the j-th sample in a batch (batch size is B) being class $c$. Correspondingly, \(g_{ij}(c)\) is the ground truth for voxel n at the j-th sample in a batch being class $c$.

\subsubsection{batch dice loss}
\label{sssec:b-dice-loss}
The dice loss turns pixel-wise labeling problem into minimizing class-level distribution distance\cite{salehi2017tversky}, thereby, dice loss is unaffected by the spatial size imbalance problem. But the frequency of each class of contributing to the dice loss computation varies greatly resulting from the high variance of the average occupied scans of each OAR. Batch dice loss\cite{kodym2018segmentation}can significantly alleviate the length imbalance problem by taking a batch of segmentation predictions maps as a single one.


To illustrate the benefits of batch dice loss, we will take hypophysis and right lung as examples. In our training set, the total number of scans is 19113, and 3094 of them has right lung annotated but only 85 of them has hypophysis annotated. If we use the original dice loss, the expected frequency that right lung participate in the loss computation is $\frac{3094}{19113} = 0.1619$, while that of hypophysis is $\frac{85}{19113} = 0.0044$. On the contrary, using our batch dice loss, if the batch size is set as 16, the expected frequency that right lung participate in the loss computation is $\min_{}(\frac{3094*16}{19113}, 1) = 1$ and this expected frequency of hypophysis is $\min_{}(\frac{85*16}{19113}, 1) = 0.0712$. In this case, the frequency ratio goes down to 14.05 from 36.4, so the length imbalance problem could be greatly mitigated by using batch dice loss.

\subsubsection{spatially balanced focal loss}
\label{sssec:sb-focal-loss}
Focal loss\cite{lin2017focal} would force model to learn more about poorly classified voxels/pixels. More importantly, focal loss would help dice loss deal with small-volume organs. It is because the gradient of dice loss $D$ regarding the prediction for voxel $i$, which can be written as $\frac{\partial D}{\partial p_{j}}=2 \times \frac{g_{j} \sum_{i}^{N}\left(p_{i}+g_{i}\right)-\sum_{i}^{N} p_{i} g_{i}}{\left(\sum_{i}^{N} p_{i}+\sum_{i}^{N} g_{i}\right)^{2}}$, would be very small if the sum of prediction probabilities is much higher than the number of foreground voxels/pixles. Thus, focal loss will potentially fasten the training of small organs. Our proposed spatially balance focal loss would foucs on the hard voxles/pixels from small-volume organs. It can be written as
\vspace{-3mm}
\begin{equation}
\mathcal{L}_{sb\_focal}  = - \frac{1}{N}  \sum\limits_{c=0}^{C-1} w_c [\sum\limits_{j=1}^B  \sum\limits_{i=1}^N  g_{ij}(c)(1-p_{ij}(c))^2\ln(p_n(c))] \end{equation}

where \(w_c=1/(\sum\limits_{j=1}^B  \sum\limits_{i=1}^N  g_{ij}(c)) \) is the inverse of individual organ volumes, designed to handle organ size imbalances in x-y plane.

\subsection{Handling partial annotations}
\label{ssec:partial-ann}
Our training dataset is composed with three sub-datasets and there exits an annotation problem. For example, liver shows in both abdomen CT images and thorax CT images, but liver is only annotated in abdomen CT images. We define such problem as partial annotations. To deal with such inconsistent annotation problem, based on the prior information regarding the source and region of the input CT scans, our re-calibration operation can be formulated as

\vspace{-5mm}
\begin{equation}
\widehat{p}_n^{rs}(c) ={p_n^{rs}(c)*w^{rs}(c)}, \:\:\:\:\:c = 0,1,...,C-2
\end{equation}

Here \(r \in \{0,1,2\}\) represents our three data sources. We employ a mask vector \(w^{rs}\) for the s-th CT image. That is for voxel n, \(w_n^{rs}(c)=0\) if annotation of organ $c$ is missed in data source $r$, and 1 otherwise. For the background, \(\widehat{p}_n^{rs}(c)={1- \sum\limits_{i=0}^{C-2} p_n^{rs}(i)*w^{rs}(i) }\) implies that we transport the predicted probability of missing-annotated organs into the background probability.

\subsection{Implementation details and performance evaluation}
\label{sec:detail}
In our experiments, a seven-fold cross validation was performed to demonstrate the performance of AttentionAnatomy. Batch size was set to 16 and Adam was used as the optimizer during training. The first 20 epochs was for pre-training the classification branch, where $\gamma$ was set to be 1, $\alpha$ and $\beta$ are 0, and learning rate was 0.001. From 20 to 50 epochs, $\alpha$, $\gamma$ and $\beta$ were all set to 1. From 50 to 70 epochs, spatially balanced focal loss was not be involved into total loss and learning rate decreased to 0.0005. During the fine-tune phrase, we introduced 2d elastic transform for data augmentation, reduced learning rate to 0.0001 and set $\alpha$, $\beta$, $\gamma$ as 1, 0, 0 respectively. Additionally, we use DSC and  95\% Hausdorff distance\cite{huttenlocher1993comparing} as the final evaluation metric.

\section{Results}
\label{sec:results}
In our experiment, the baseline results were generated from a Vanilla Unet. The encoder and segmentation branch of our proposed model are exactly the same as the baseline model. Thus, the GPU computation of our model is only 1.02\% larger than our baseline model.

\begin{figure}
\centering
\centerline{\includegraphics[width=9cm]{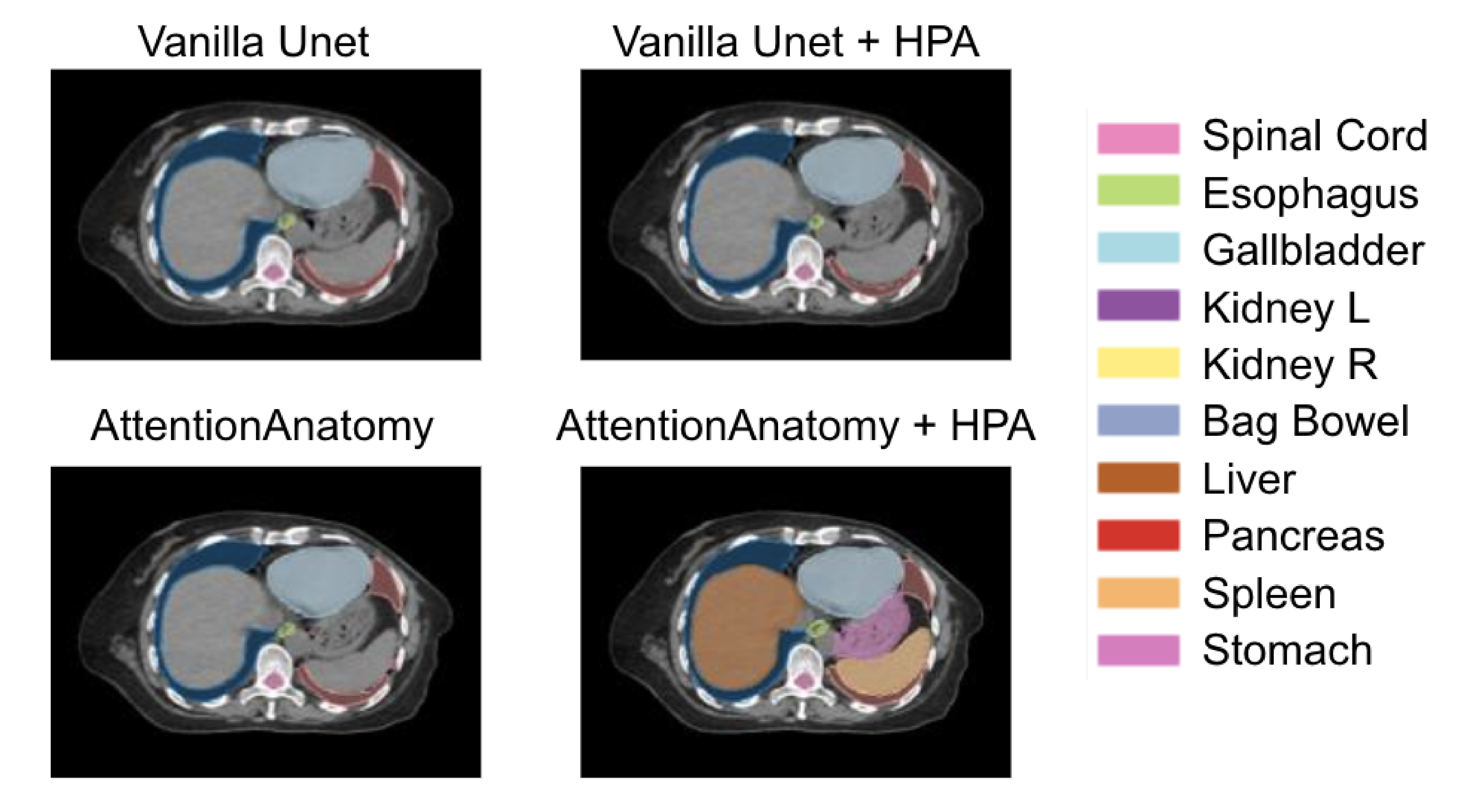}}
\caption{Prediction visualization of Vanilla Unet, Vanilla + HPA, AttentionAnatomy and AttentionAnatomy + HPA. HPA is short for handling partial-annotation problem.}
\label{fig:visualization}
\end{figure}

Here we can see from Table.\ref{table:dice-results}, AttentionAnatomy had a significant improvement over baseline. DSC is increased by an average of 2.84\% and the 95\% Hausdorff distance is lowered from an average of 11.42 mm to 9.33 mm. We can also learn from Table.\ref{table:dice-results} that the method of handling partial -annotation problem will be more beneficial if applied in our proposed framework.  It is because our proposed attention module would provide the posterior information that what OARs the input CT may contain. With the restriction of such posterior information, the probability re-calibration results would be more convincing.
Fig.\ref{fig:visualization} strongly supports the numerical results in Table.\ref{table:dice-results} by visualizing the predicted segmentation of partially-annotated region, generated from different models and settings.

\begin{center}
\bottomcaption{DSC comparison between baseline model and our proposed model. L is short for left and R is short for right. SMG is short for submandibular gland and TMJ is short for temporomandibular joint. Att-Anatomy is short for AttentionAnatomy.
\label{table:dice-results}}
\tablefirsthead{\toprule OARs & Vanilla Unet & Vanilla Unet + HPA & Att-Anatomy & Att-Anatomy + HPA \\ \midrule}
\tabletail{\midrule \multicolumn{3}{r}{{\bfseries Continued on next column}} \\ \midrule}
\tablehead{\multicolumn{3}{l} {{\bfseries  Continued from previous column}} \\
    \toprule
    OARs & Vanilla Unet & Vanilla Unet + HPA & Att-Anatomy & Att-Anatomy + HPA \\
    \midrule}
\tablelasttail{\toprule}
\tiny
\begin{supertabular}{l l l l l}
Brain Stem & $83.17 \pm 1.22$ & $84.21 \pm 1.36$ & $83.89 \pm 2.53$ & $\textbf{85.87 $\pm$ 1.35}$ \\
Constrictor Naris & $72.35 \pm 6.47$ & $71.82 \pm 5.72$ & $74.93 \pm 6.53$ & $\textbf{77.14 $\pm$ 4.31}$ \\
Ear L & $74.04 \pm 3.58$ & $\textbf{78.78 $\pm$ 3.85}$ & $78.14 \pm 3.82$ & $77.20 \pm 6.34$ \\
Ear R & $77.57 \pm 4.17$ & $81.04 \pm 3.55$ & $80.61 \pm 3.81$ & $\textbf{82.13 $\pm$ 4.18}$ \\
Eye L & $87.28 \pm 4.09$ & $88.66 \pm 4.17$ & $89.54 \pm 1.89$ & $\textbf{90.10 $\pm$ 2.00}$ \\
Eye R & $88.07 \pm 0.94$ & $88.02 \pm 1.44$ & $88.94 \pm 1.56$ & $\textbf{90.41 $\pm$ 0.71}$ \\
Hypophysis & $61.71 \pm 6.71$ & $61.17 \pm 6.47$ & $58.68 \pm 11.27$ & $\textbf{69.07 $\pm$ 8.20}$ \\
Larynx & $89.99 \pm 0.58$ & $90.45 \pm 0.83$ & $91.01 \pm 1.29$ & $\textbf{92.20 $\pm$ 0.68}$ \\
Mandible & $93.04 \pm 1.06$ & $94.04 \pm 0.75$ & $\textbf{94.27 $\pm$ 0.56}$ & $94.05 \pm 1.08$ \\
Oral Cavity & $89.62 \pm 3.52$ & $89.83 \pm 2.80$ & $\textbf{90.50 $\pm$ 2.58}$ & $89.37 \pm 2.63$ \\
Parotid L & $75.41 \pm 4.23$ & $76.37 \pm 3.46$ & $79.33 \pm 4.20$ & $\textbf{83.81 $\pm$ 3.48}$ \\
Parotid R & $78.54 \pm 6.98$ & $77.99 \pm 8.61$ & $79.85 \pm 4.79$ & $\textbf{80.81 $\pm$ 4.10}$ \\
Smg L & $73.16 \pm 6.41$ & $70.93 \pm 6.83$ & $72.72 \pm 10.81$ & $\textbf{77.93 $\pm$ 5.37}$ \\
Smg R & $67.11 \pm 10.98$ & $68.09 \pm 11.45$ & $71.94 \pm 10.07$ & $\textbf{77.24 $\pm$ 7.10}$ \\
Spinal Cord & $65.61 \pm 31.62$ & $87.90 \pm 5.49$ & $88.28 \pm 5.05$ & $\textbf{89.13 $\pm$ 5.37}$ \\
Sublingual Gland & $37.26 \pm 17.34$ & $42.05 \pm 10.87$ & $42.14 \pm 11.55$ & $\textbf{46.73 $\pm$ 12.67}$ \\
Temporal Lobe L & $87.30 \pm 3.67$ & $\textbf{88.78 $\pm$ 3.45}$ & $88.23 \pm 2.69$ & $87.03 \pm 3.45$ \\
Temporal Lobe R & $86.99 \pm 4.62$ & $87.65 \pm 4.34$ & $\textbf{88.58 $\pm$ 2.90}$ & $86.09 \pm 4.53$ \\
TMJ L & $78.48 \pm 6.05$ & $74.34 \pm 13.98$ & $82.76 \pm 4.90$ & $\textbf{84.94 $\pm$ 3.27}$ \\
TMJ R & $77.42 \pm 6.42$ & $83.06 \pm 3.48$ & $79.67 \pm 15.08$ & $\textbf{88.65 $\pm$ 1.81}$ \\
Trachea & $88.98 \pm 2.34$ & $90.34 \pm 1.96$ & $90.20 \pm 1.66$ & $\textbf{91.53 $\pm$ 1.47}$ \\
Heart & $90.46 \pm 4.01$ & $90.47 \pm 4.12$ & $\textbf{91.60 $\pm$ 2.52}$ & $90.62 \pm 2.51$ \\
Lung L & $96.86 \pm 1.89$ & $96.81 \pm 2.16$ & $97.03 \pm 1.24$ & $\textbf{97.70 $\pm$ 0.60}$ \\
Lung R & $96.84 \pm 1.69$ & $96.57 \pm 2.45$ & $97.23 \pm 0.54$ & $\textbf{97.57 $\pm$ 0.49}$ \\
Eso & $73.88 \pm 1.85$ & $72.92 \pm 3.82$ & $74.10 \pm 2.81$ & $\textbf{79.38 $\pm$ 2.07}$ \\
Gallbladder & $61.03 \pm 21.47$ & $66.28 \pm 19.65$ & $62.01 \pm 23.28$ & $\textbf{68.81 $\pm$ 25.14}$ \\
Kidney L & $93.38 \pm 1.14$ & $\textbf{94.09 $\pm$ 1.46}$ & $92.24 \pm 4.59$ & $94.04 \pm 1.66$ \\
Kidney R & $93.65 \pm 1.55$ & $93.88 \pm 2.04$ & $92.97 \pm 3.09$ & $\textbf{94.87 $\pm$ 1.79}$ \\
Bag Bowel & $\textbf{81.49 $\pm$ 2.93}$ & $80.20 \pm 3.11$ & $78.99 \pm 3.92$ & $80.94 \pm 3.59$ \\
Liver & $91.28 \pm 2.70$ & $90.57 \pm 3.63$ & $77.29 \pm 18.58$ & $\textbf{93.89 $\pm$ 1.84}$ \\
Pancreas & $62.83 \pm 14.31$ & $59.78 \pm 16.81$ & $54.96 \pm 19.24$ & $\textbf{64.77 $\pm$ 13.54}$ \\
Spleen & $87.53 \pm 8.89$ & $86.61 \pm 10.19$ & $79.82 \pm 13.57$ & $\textbf{90.29 $\pm$ 5.92}$ \\
Stomach & $62.95 \pm 16.56$ & $60.73 \pm 17.05$ & $47.41 \pm 12.53$ & $\textbf{63.97 $\pm$ 17.58}$ \\
\hline
Average & $79.55$& $80.74$& $80.00$& $\textbf{83.58}$\\ \hline
\end{supertabular}
\end{center}

\section{CONCLUSIONS}
\label{sec:conclusions}

Deep learning based OARs delineation solutions have proved comparable to expert standard. However, most of these researches only focused on one particular region. In this work, we proposed a light, flexible and more clinical applicable end-to-end framework aiming to segment whole-body OARs. What's more, we incorporate an attention module connecting segmentation and classification branches to guide the segmentation predication, and a re-calibration methods to tackle the partial-annotation problem.

\bibliography{root.bbl}

\begin{thebibliography}{1}

\bibitem{milletari2016v}
Fausto Milletari, Nassir Navab, and Seyed-Ahmad Ahmadi,
\newblock ``V-net: Fully convolutional neural networks for volumetric medical
  image segmentation,''
\newblock in {\em 2016 Fourth International Conference on 3D Vision (3DV)}.
  IEEE, 2016, pp. 565--571.

\bibitem{ronneberger2015u}
Olaf Ronneberger, Philipp Fischer, and Thomas Brox,
\newblock ``U-net: Convolutional networks for biomedical image segmentation,''
\newblock in {\em International Conference on Medical image computing and
  computer-assisted intervention}. Springer, 2015, pp. 234--241.

\bibitem{tang2019clinically}
Hao Tang, Xuming Chen, Yang Liu, Zhipeng Lu, Junhua You, Mingzhou Yang, Shengyu
  Yao, Guoqi Zhao, Yi~Xu, Tingfeng Chen, et~al.,
\newblock ``Clinically applicable deep learning framework for organs at risk
  delineation in ct images,''
\newblock {\em Nature Machine Intelligence}, pp. 1--12, 2019.

\bibitem{zhu2019anatomynet}
Wentao Zhu, Yufang Huang, Liang Zeng, Xuming Chen, Yong Liu, Zhen Qian, Nan Du,
  Wei Fan, and Xiaohui Xie,
\newblock ``Anatomynet: Deep learning for fast and fully automated whole-volume
  segmentation of head and neck anatomy,''
\newblock {\em Medical physics}, vol. 46, no. 2, pp. 576--589, 2019.

\bibitem{salehi2017tversky}
Seyed Sadegh~Mohseni Salehi, Deniz Erdogmus, and Ali Gholipour,
\newblock ``Tversky loss function for image segmentation using 3d fully
  convolutional deep networks,''
\newblock in {\em International Workshop on Machine Learning in Medical
  Imaging}. Springer, 2017, pp. 379--387.

\bibitem{kodym2018segmentation}
Old{\v{r}}ich Kodym, Michal {\v{S}}pan{\v{e}}l, and Adam Herout,
\newblock ``Segmentation of head and neck organs at risk using cnn with batch
  dice loss,''
\newblock in {\em German Conference on Pattern Recognition}. Springer, 2018,
  pp. 105--114.

\bibitem{lin2017focal}
Tsung-Yi Lin, Priya Goyal, Ross Girshick, Kaiming He, and Piotr Doll{\'a}r,
\newblock ``Focal loss for dense object detection,''
\newblock in {\em Proceedings of the IEEE international conference on computer
  vision}, 2017, pp. 2980--2988.

\bibitem{huttenlocher1993comparing}
Daniel~P Huttenlocher, Gregory~A Klanderman, and William~J Rucklidge,
\newblock ``Comparing images using the hausdorff distance,''
\newblock {\em IEEE Transactions on pattern analysis and machine intelligence},
  vol. 15, no. 9, pp. 850--863, 1993.

\end{thebibliography}

\end{document}